# DYNAMIC ACCESSIBILITY USING BIG DATA: THE ROLE OF THE CHANGING CONDITIONS OF NETWORK CONGESTION AND DESTINATION ATTRACTIVENESS

MOYA-GÓMEZ, BORJA (*)

Transport, Infrastructure and Territory Research Group (t-GIS)

Human Geography Department

Faculty of Geography and History

Universidad Complutense de Madrid (UCM)

c/ Profesor Aranguren S/N, 28040 Madrid, Spain

Phone: +34 91 394 57 51

e-mail: bmoyagomez@ucm.es

SALAS-OLMEDO, MARÍA HENAR

Transport, Infrastructure and Territory Research Group (t-GIS)

Human Geography Department

Faculty of Geography and History

Universidad Complutense de Madrid (UCM)

c/ Profesor Aranguren S/N, 28040 Madrid, Spain

Phone: +34 91 394 59 49

e-mail: msalas01@ucm.es

GARCÍA-PALOMARES, JUAN CARLOS

Transport, Infrastructure and Territory Research Group (t-GIS)

Human Geography Department

Faculty of Geography and History

Universidad Complutense de Madrid (UCM)

c/ Profesor Aranguren S/N, 28040 Madrid, Spain

Phone: +34 91 394 59 52

e-mail: jcgarcia@ghis.ucm.es

GUTIÉRREZ, JAVIER

Transport, Infrastructure and Territory Research Group (t-GIS)

Human Geography Department

Faculty of Geography and History

Universidad Complutense de Madrid (UCM)

c/ Profesor Aranguren S/N, 28040 Madrid, Spain

Phone: +34 91 394 59 49

e-mail: javiergutierrez@ghis.ucm.es

* Corresponding author

**ABSTRACT**

Accessibility is essentially a dynamic concept. However, most studies on urban accessibility take a static approach, overlooking the fact that accessibility conditions change dramatically throughout the day. Due to their high spatial and temporal resolution, the new data sources (Big Data) offer new possibilities for the study of accessibility. The aim of this paper is to analyse urban accessibility considering its two components –the performance of the transport network and the attractiveness of the destinations– using a dynamic approach using data from TomTom and Twitter respectively. This allows us to obtain profiles that highlight the daily variations in accessibility in the city of Madrid, and identify the influence of congestion and the changes in location of the population. These profiles reveal significant variations according to transport zones. Each transport zone has its own accessibility profile, and thus its own specific problems, which require solutions that are also specific.

# 1 INTRODUCTION

Accessibility is a key concept in urban and regional planning for its capacity to link the activities of people and businesses to the possibilities of reaching them effectively. Accessibility therefore simplifies the relationship between land use and transport, and models the interaction between both systems. Interest in the relationship between transport systems and spatial interaction has grown exponentially, and for more than five decades accessibility analysis has played a key role in the agenda of regional and transport research (Reggiani and Martín 2011).

Accessibility is a dynamic attribute of locations that varies over time due to changes in the transport network and in the attractiveness of destinations for certain activities. One of the challenges when studying accessibility is to improve the method of introducing the spatial-temporal dimension, and particularly the analysis of daily changes (Geurs and van Wee 2004; Geurs et al. 2015; van Wee 2016), an issue that has scarcely been addressed until now due to the limitations of traditional data sources. Some works have used these data sources to analyse the effect of congestion on accessibility, considering solely extreme situations (peak and valley times) based on the network (for example, Vandenbulcke et al. 2009; Yiannakoulias et al. 2013), but not the temporal changes in the performance of the infrastructures that occur throughout the day.

Today's new big data sources offer exciting opportunities for the dynamic analysis of accessibility. The information on transport networks has improved conclusively in recent years thanks to the emergence of big data generated by social media, smartphones, Satnav and other technologies (van Wee 2016). Navigation companies such as TomTom, NavTeq, Inrix and more; websites like Here, Bing Maps, Google Maps-Google Transit; collaborative projects like Open-Street-Map; and the public availability of Transit Feed Specification (GTFS) data from transit authorities, among others, open up a growing field of research on time-of-day variations in private and public transit accessibility (Geurs et al. 2016). These companies and institutions have increasingly detailed systems with plentiful information on the features of roads and public transport networks, and their databases include information on speed variations on the roads and the frequencies of passage in public transport networks, all of which contribute a more efficient and dynamic vision to intraurban accessibility studies.

There is very little research using the new sources in studies on travel times and accessibility. A first group of papers employ data obtained from GPS devices. Møller-Jensen et al. (2012) used GPS logs to calculate speeds, congestion levels and accessibility conditions at three times of day (morning, midday, evening,) in the city of Accra. Dewulf et al. (2015) took Floating Car Data (FCD) from the Be-Mobile system to calculate car travel times. Be-Mobile provides the geolocated positions of 400,000 vehicles equipped with track and trace devices. The travel time measures are further aggregated to produce a generic travel time for peak and off-peak periods. Elsewhere, Owen and Levinson (2015) calculated car accessibility dynamically from data recorded by loop detectors and GPS data.

Other works have studied information from web services to calculate travel times between origins and destinations. Martin et al. (2002; 2008) incorporated public transport timetable data from a web service to analyse accessibility to hospitals in England. Páez et al. (2013) developed a web-based accessibility instrument using Google Maps API to retrieve information about local amenities (e.g. groceries, restaurants, fitness centres, banks and others) and estimate accessibility by car, walking and cycling. Farber et al. (2014) used General Transit Feed Specification (GTFS) data to calculate dynamic accessibility to food stores by transit.

Boisjoly and El-Geneidy (2016) calculated transit accessibility to jobs, accounting for fluctuations in job availability (mobility survey data) and transit service (GTFS data) throughout the day. Jäppinen et al. (2013) used transport information interfaces for Greater Helsinki to analyse improvements in public transport times after a complementary bike sharing system was added to the network.

Of particular interest for this work are the dynamic data from companies that offer daily Speed Profiles with a high spatial and temporal resolution. These companies use crowd-sourcing, Big Data analytics and location technologies to obtain real-time traffic and speed data that leverage current, historical and predictive traffic information across the roadway network. The best-known of these companies is Google, although Inrix and TomTom also provide this type of data. The works published using this type of data sources include particularly those from Toronto by Sweet et al. (2014, 2015), who studied the impact of congestion on accessibility and its consequences on company localisation using historic speed data for motorways and major arterial networks provided by Inrix. Elsewhere, Moya-Gómez and García-Palomares (2015) applied data supplied by the TomTom navigation company –specifically the "Speed Profiles" product– to create dynamic maps that reveal the impact of congestion on daily accessibility in the metropolitan area of Madrid. "Speed Profiles" shows the speed on each section of the road network every five minutes. These are historical data obtained from different devices, including the company's own navigators and mobile phone GPS. This new source has so far been underexploited in the study of accessibility.

In addition to network performance, the study of daily accessibility also needs to incorporate the effect of variations in the attractiveness of destinations for certain desired activities. In static accessibility analyses, destination attractiveness is measured through variables such as population or employment. However, dynamic analyses take into account that the destinations' attractiveness changes throughout the day. The population distribution in the city varies at different times of day depending on the type of activities that predominate in each time band (for example, work in the morning, shopping and leisure in mid-afternoon), and can be analysed from the population's digital footprint every hour of the day and at each point in the city. A dynamic focus must therefore be included in the accessibility analysis, based on activities (where the population is at each time of the day) as a proxy for the destinations' attractiveness. Traditional data sources (censuses) offer information on the spatial distribution of the population at night (place of residence) but not on their location throughout the day, whereas the new data sources allow a dynamic approach to population distribution in the city. Mobile phone logs –so-called CDR (Call Detail Records)– have been used to map the density of mobile phone activity at different times of the day as an indicator of spatial-temporal changes in the population density in the city (Ratti et al. 2006; Reades et al. 2009). As the density of calls varies in different time bands and reflects the changes in population densities, each area of the city has its own signature; that is to say, a time profile of mobile phone use, which is very frequent in areas of activity in the central hours of the day, whereas in residential areas it is higher in the afternoon and early evening (Reades et al. 2009; Louail et al. 2014; Grauwin et al. 2015).

Social networks also reflect the rhythms of the city. The most widely used data from social networks in urban studies come from Twitter (see Murthy 2013), due to its considerable reach and the fact that the tweets can be downloaded free from the Internet. The densities of tweets sent from each area of the city reveal the population densities (Jiang et al. 2016). Maps of tweet density can be obtained according to the age, gender and ethnic group of the tweeter, if this information can be inferred from the user identifier (Longley et al. 2015). One approach to the analysis of the daily changes in the population distribution in the city is to map the spatial distribution of the tweets according to the time of day (Ciuccarelli et al. 2014). In addition to the official statistics that show the population's place of residence, the spatial-temporal analysis of tweets is now making it possible to move beyond night-time geographies of residence to see how they compare with daytime activity patterns (Longley et al. 2015).

The spatial-temporal analysis of tweets allows researchers to track users and deduce their mobility patterns (Wu et al. 2014), and reveals the spaces where different population groups converge, either based on income bands (Netto et al. 2015) or race (Shelton et al. 2015). The reliability of Twitter data in mobility studies has been validated in the work of Lenormand et al. (2014), who compared the data from Twitter, mobile telephony and official data (censuses), and concluded that the three information sources offer comparable results. However, no research has been found in which these new data sources (mobile telephony, social networks and others) have been used as a proxy for the variability in destinations' attractiveness in the study of dynamic accessibility.

The literature review uncovers very few works that have applied these new data sources to the study of accessibility, and very few that have calculated accessibility in a dynamic way. Those that do, look at the temporal variation in only one component of accessibility, namely network performance. The aim of this paper is to analyse the variations in daily accessibility integrating the time variability of access times and the attractiveness of the destinations. We used the TomTom "Speed Profiles" product to study the variation in travel times (the effect of congestion), which includes the speeds on each roadway section every five minutes. The daily variation in the destinations' attractiveness was studied using geolocated Twitter data. Each message contains the time and the geographic coordinates of the place from which the tweet is sent, along with the user identifier. This enables a map to be created showing the distribution of Twitter users for each transport zone in the study area every 15 minutes, for use as a proxy for the attractiveness of each zone at each time of day. These data sources are used to make a dynamic study of accessibility, and analyse the influence of each accessibility component (network performance and the attractiveness of the destinations for activities) in each transport zone and at each time of day. The study area is the city of Madrid.

The paper contributes to the literature in several ways. To the best of our knowledge, this is the first time an accessibility analysis has been undertaken using new data sources with global coverage (TomTom and Twitter) considering the two components of accessibility dynamically, namely travel time and destination attractiveness. Accessibility is calculated every 15 minutes, which provides a sequence of high-resolution time maps, instead of several "fixed photos", which are then animated in the form of videos, and the time profile (the signature) of each transport zone. Finally, assessment scenarios are built to give an insight into the influence of both accessibility components as they change throughout the course of the day.

The article is structured as follows. After this introduction, section 2 describes the data and their pre-processing to obtain the dynamic variables necessary for the calculation of the accessibility indicator. Section 3 presents the accessibility indicator and the assessment scenarios used to analyse dynamic accessibility and differentiate the effects of network congestion and destination attractiveness. Section 4 contains the results of this research, and finally section 5 presents the main conclusions.

## 2 STUDY CONTEXT

Madrid was chosen as the study area for testing the proposed methodology. The metropolitan area of Madrid has a total of 6 million inhabitants and 2.5 million jobs, but the distribution of the population and employment inside the city is very uneven. The city centre is home to 55% of the population, but 65% of the city's jobs. Other differences include a marked contrast between the north of the city –where a substantial proportion of the economic activities have been displaced– and the south, with a distinctly residential nature. As a result of this uneven distribution of population and employment, there is a predominance of flows from the suburbs to the centre and from the south to the north in the early hours of the morning, and from the centre to the suburbs and from north to south in the evening, which has direct consequences on road congestion and accessibility. The inbound radial motorways and the orbital motorways in the north-south direction are congested at the morning peak time, whereas the opposite occurs at the afternoon peak time.

## 3 DATA DESCRIPTION AND DATA PRE-PROCESSING

Two datasets are involved in the accessibility measures presented in this paper: the dataset for the distribution of the attraction factor, and the road network dataset that allows the computation of OD travel times. As our aim is to analyse dynamic accessibility (across the metropolitan area of Madrid), both datasets must have fine spatial and temporal resolutions.

### 3.1 TomTom Speed Profiles and minimum travel time calculations

This study uses the March 2013 version of TomTom® for the road network, which contains data on historical Speed Profiles for the years 2011 and 2012 obtained from the average journey times reported from users' navigation devices. As the original network is very detailed (it includes accesses to car parks, pedestrian streets, residential streets and country roads), arcs where not much traffic is expected have been omitted. The arcs used in the study are defined by TomTom® as ranging from 0 to 6 in the Functional Road Classification (FRC). The network covers the whole metropolitan area of Madrid and has full connectivity, with a total of 12,935 km (18,235 km one-way arcs), of which 81% have historic Speed Profiles.

Historic Speed Profiles are defined every 5 minutes as a percentage of the free-flow speed of the arc. As a result, the arc of a motorway and a city street may both have the same speed profile but different speeds at the same instant because of their different free-flow speeds. This data structure saves on computational memory and cost and is prepared for use with the GIS software ESRI® ArcGIS. An aggregation of these speed variations every 15 minutes is shown in video 1.

We calculated the travel time per OD pair of transport zones every 15 minutes. This includes the speed variations throughout the journey. Our results were first grouped by start times, which is the only output option of the OD Cost Matrix ArcGIS tool. However, we wanted to group them by end time, i.e. how long an individual must travel in order to reach their destination zone at a certain time. We interpolated these values from the original matrices by splines using the SciPy Python library, as we knew the arrival time and travel time for each OD pair.

**3.2 Geolocated tweets and attractiveness of destinations**

We chose to use Twitter data as a proxy to estimate the attraction factor of different transport zones in the study area. Table 1 shows the main figures in the dataset of geolocated tweets published between January 2012 and December 2013 in our study area, which is the basis of our research.

Table 1. Geolocated tweets in the Region of Madrid, main figures

|  | TOTAL |
|---|---|
| Tweets (all year) | 12,408,065 |
| Tweets (published on a Tuesday, Wednesday or Thursday) | 5,546,200 |
| Single users in each zone at 15-minute intervals | 692,117 |

The number of active users in each transport zone at each time of day was calculated using a Geographic Information System (ArcGIS 10.3). The data was treated to convert single tweets into an attraction factor that varies spatially and temporally by first selecting the tweets corresponding to typical working days (i.e. Tuesday to Thursday)1, and second, computing a joint spatial (transport zones2) and temporal (15-minute period) aggregation in order to obtain the number of single users in each zone at each time of the day (see video 2).

Figure 1 shows the wide variation in active Twitter users throughout the day, with a clear minimum in the central hours of the night and a maximum between 8 pm and 10 pm. The night hours when the population is not engaged in any activity are of no interest from the point of view of accessibility analysis. The period between midnight and 7 am was thus discarded, which in Madrid tends to correspond to the period of night-time rest. Between 7 am and midnight there is an uneven distribution of active tweeters, although the size of the population remains the same (approximately); the individuals simply move from some places to others within the metropolitan area. We therefore opted to normalise the data on active tweeters in units per 100,000 to obtain a proxy for the spatial distribution of the population according to the place they are tweeting from throughout the day, in order to estimate the attractiveness of each transport zone in the accessibility analysis. The greater the size of the population in a transport zone, the greater its attractiveness at that time of day.

---

[1] Longley et al. (2015) report that these three days of the week have a very similar tweet profile, and represent the average working day, while Monday and Friday have specific profiles influenced by the proximity of Sunday and Saturday respectively.
[2] Transport zones correspond to homogeneous land-use areas. There are 1,171 transport zones in the study area.

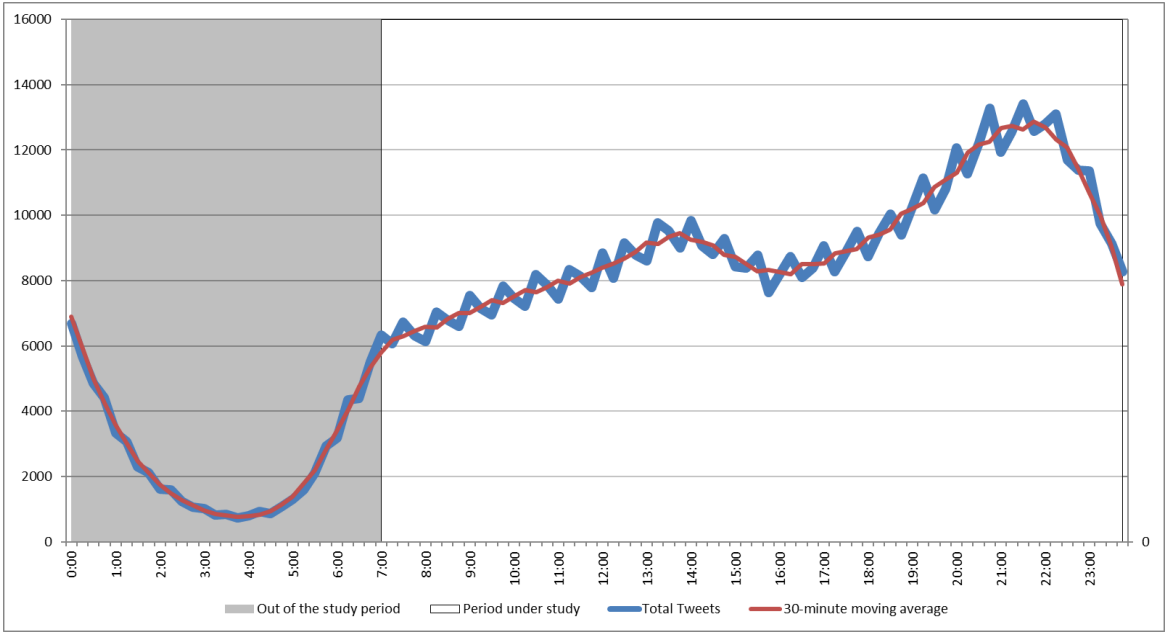

Figure 1: Number of active tweeters throughout the day in the metropolitan area of Madrid

## 4 METHODS

### 4.1 Accessibility calculations

The dynamic approach proposed in this paper involves calculating the accessibility of each transport zone every 15 minutes based on the spatial distribution of the population (normalised number of tweeters) in that quarter hour, and the travel times between each OD pair according to the arrival time at the destination transport zone. This is done by using the potential accessibility indicator, which shows the spatial interaction between a specific origin and all the destinations taking into consideration the degree of attractiveness of each destination, the cost of travel and the distance decay, i.e., how fast the interaction drops as the transport cost increases (Reggiani et al., 2011). Since it is generally agreed that exponential functions are more appropriate for analysing short distance interactions such as those occurring within urban areas (Bruno and Genovese, 2012), we computed the potential accessibility using an exponential function according to the following formula:

$$P_{it} = \sum_{j=1}^{n} M_{jt} \cdot e^{\alpha \cdot C^{*}_{ijt}} \qquad (1)$$

Where $P_{it}$ is the potential accessibility of transport zone *i* at time *t*, $M_{jt}$ is the normalized number of tweets in transport zone j at time t, $C^{*}_{ijt}$ is the travel time through the road network between transport zones *i* and *j* at time *t*, and α is a parameter indicating the distance decay. In this case, we calibrated α using the Hyman algorithm (Ortúzar and Willumsen, 2011, 192) with the 2004 Region of Madrid Mobility Survey. The resulting parameter was -0.12957849.

To determine the self-potential, the internal transport zone times were calculated as the average minimum travel time between 10% of the randomly chosen network junctions to their centroid zone.

Finally, we added half the travel time of the origin/destination zone to each journey. So $C^{*}_{ijt}$ was calculated according to:

$$C^*_{ijt} = \frac{1}{2}C_{ii} + C_{ijt} + \frac{1}{2}C_{jj} \qquad (2)$$

To isolate the effect of each accessibility component (ease of access via the network and destination attractiveness) on dynamic accessibility, this indicator was calculated for each scenario described in subsection 4.2. Given that the approach taken in this research requires numerous accessibility calculations (depending on the scenarios and times of day), this indicator was integrated in a single ArcGIS toolbox in order to facilitate its computation.

### 4.2 Scenarios

Four scenarios were considered for the analysis of dynamic accessibility and the influence of its different components (travel time and destination attractiveness):

a) Reference scenario.- Accessibility is calculated based on average travel times and the average spatial distribution of the tweeters throughout the day. This is therefore a static scenario that is taken as a reference to assess the changes that occur throughout the day based on the temporal variations in congestion and the spatial distribution of the population within the metropolitan area.

b) Dynamic accessibility scenario.- Accessibility is calculated every 15 minutes taking into account the variability in congestion and the spatial distribution of the tweeters.

c) Dynamic congestion scenario.- Accessibility is calculated every 15 minutes considering the variation in congestion levels, while the population distribution remains static (average spatial distribution of the tweeters throughout the day). This makes it possible to isolate the effect of the variation in congestion levels on accessibility.

d) Dynamic attractiveness scenario.- Accessibility is calculated every 15 minutes, but in this case only the variation in the population distribution is considered dynamic, whereas congestion remains static (average travel times).

The dynamic accessibility analysis is compared with the reference scenario to identify the differences (according to the times of day) produced by the dynamic approach compared to the static approach. The dynamic congestion and attractiveness scenarios highlight the influence of each of these two components on dynamic accessibility.

### 5 RESULTS

Figure 2 shows the temporal pattern of average accessibility values for the whole of the metropolitan area of Madrid according to journey arrival times and scenarios. The reference scenario has an average value of around 6,200 potential units and represents the average accessibility throughout the day. The curve for the dynamic accessibility scenario is substantially higher than this average value in the early hours of the morning (between 7 and 7:30 am) and from 6:30 pm, revealing a higher than average accessibility in this time band. In contrast, it is clearly below the average for the static scenario between 7:30 am and 6:30 pm, and particularly around 8:30 and 9:30 am and between 3 and 6 pm.

The pattern of dynamic accessibility depends on the joint action of the changes in congestion and population distribution throughout the day. The curves for the dynamic congestion and attractiveness scenarios generally show opposing behaviours. If the population distribution remains fixed (dynamic congestion scenario), two clear times with lower accessibility can be observed, corresponding to the morning and afternoon peaks (high levels of congestion due to travel to and from work), whereas after 9 pm accessibility is maximum (free flow). When congestion levels remain fixed (dynamic attractiveness scenario), accessibility is above average until 7:30 pm (except at lunchtime), when the population is concentrated in the centre (the most easily accessible space), but it falls below average at the end of the day when the population tends to disperse around residential areas in the suburbs (less accessible than the centre).

Therefore, the dynamics for congestion and destination attractiveness represent opposing forces that tend to offset each other. For example, in the early hours of the morning the population is concentrated in the centre (which increases accessibility), but congestion levels are high (which reduces accessibility). This means that the dynamic accessibility curve is almost always located between the curves for the dynamic congestion and attractiveness scenarios, although the fact that it is located closer to the first than the second indicates that the variation in congestion outweighs the variation in destination attractiveness. In fact, the lowest dynamic accessibility values occur at the morning and afternoon peak times, although this is when the population tends to be concentrated in the centre (see also Table 3).

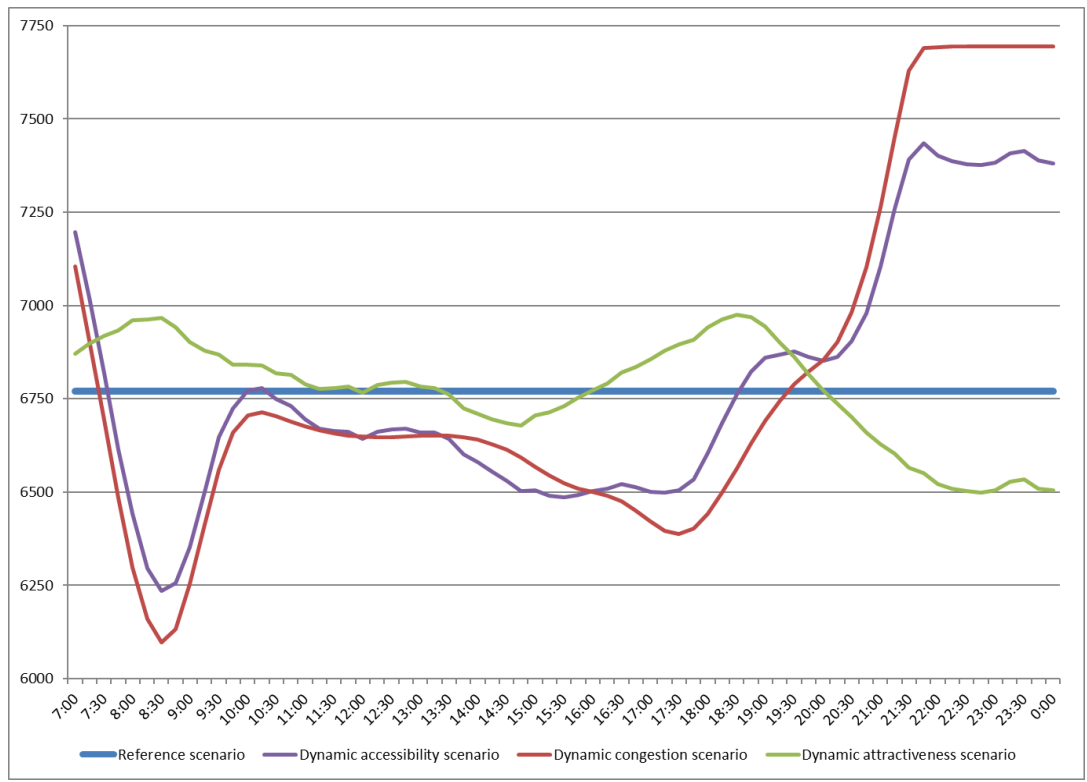

Figure 2: Changes in average accessibility every 15 mins according to scenarios

In general terms the average values of each scenario reveal the pattern of dynamic accessibility and the role played by the two dynamic variables (network congestion and destination attractiveness) in the whole of the metropolitan area of Madrid, taking the reference scenario as the element for comparison. However, this general pattern conceals major differences in the profiles of the different transport zones. As it is impossible to analyse the profiles of each zone, we have selected three that have locations and features that are representative of the internal discrepancies in the metropolitan area: *a)* a residential city in the suburban south *b)* the city centre; *c)* an area of activity in the north (the headquarters of the multinational company Telefónica); and (Figure 3). In the reference scenario, the accessibility values for the second zone (centre) are substantially higher than the average for the metropolitan area (almost 10,000 units), intermediate (5,000) in the third (north), and far below average (4,400) in the first (south). The curves for temporal changes in dynamic accessibility are very different:

> a) The residential city in the metropolitan south reveals a very marked decline in accessibility due to the effects of congestion at the morning peak time (Figure 3a) when there is a predominance of journeys from the suburbs to the centre. Unlike the general profile for the metropolitan area, the dynamic population distribution at the morning and evening peak times does not tend to offset the loss of accessibility due to congestion but instead reinforces it, as the city loses population during these periods owing to its residential character. In the central hours of the day this space suffers barely any congestion, and the dynamic accessibility is similar to that of the reference scenario.

> b) The central transport zone is not affected by congestion in the early hours of the morning when there is a predominance of inbound journeys, so its dynamic accessibility is higher than the static average in this time period (Figure 3b). However, it is significantly affected by congestion in the central hours of the day and particularly at the afternoon peak time, when the population begins to leave the centre to travel to the suburbs. The curve for the dynamic attractiveness scenario is significantly higher than the reference scenario in both the morning and afternoon, as the centre contains a substantial proportion of the population at these times.

> c) Finally, the area of economic activity in the north of the city is very sensitive to congestion at the afternoon peak time (Figure 3c), coinciding with the time people leave work. Dynamic accessibility is higher than static accessibility during the morning (until 2:30 pm), which is explained by the population concentration and the alleviation of congestion in the outbound journeys. At night the dynamic accessibility is higher than the static average in all three spaces.

In short, the dynamic accessibility profiles in the three transport zones are the result of the interplay of the opposing forces of the congestion dynamic and population density. In the centre, the population concentration and the low congestion in the early hours of the morning combine to create a situation of greater accessibility. At the end of the afternoon, the high population concentration counteracts the impact of congestion. However, during the central hours of the day, the population concentration does not offset the effects of road congestion. This situation is repeated in the north, where the high population concentration compensates for the effects of congestion in the morning, but not in the afternoon peak time. In the residential area in the metropolitan south, the population dynamic is more stable than in the two previous zones, though the dynamic accessibility profile is more similar to that of the dynamic congestion scenario. The most important finding is that the population dynamic reinforces (rather than offsets) the effects of network congestion, reducing dynamic accessibility at the morning and afternoon peak times and increasing it at the end of the day.

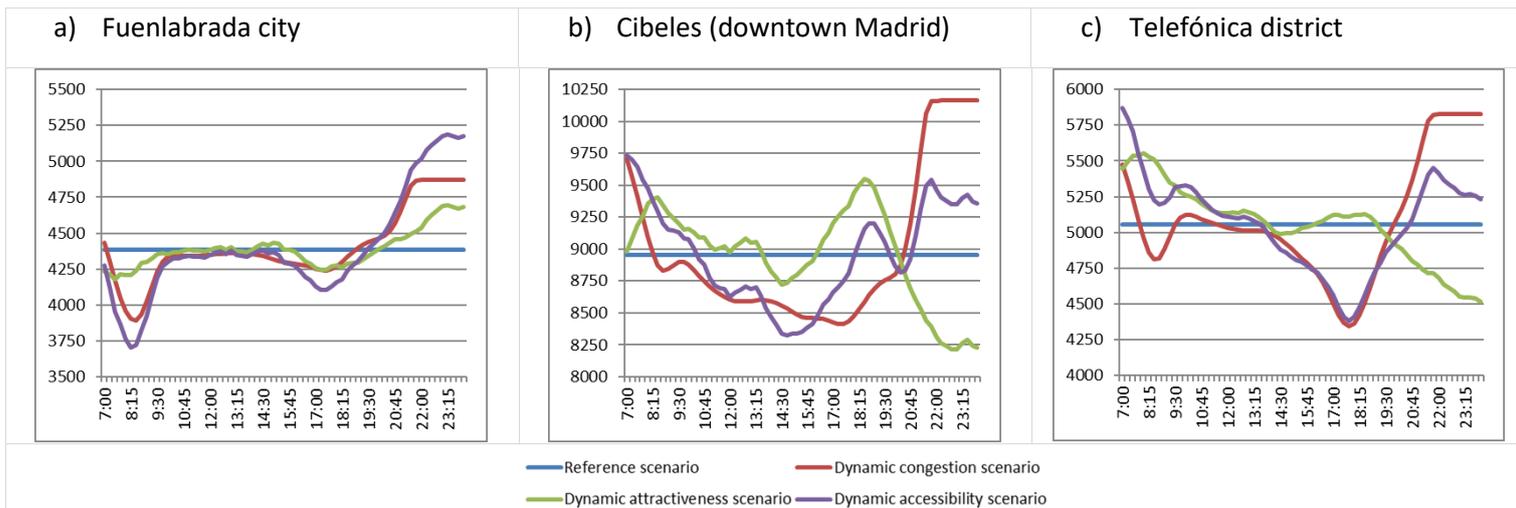

Figure 3: Profile for changes in accessibility according to scenarios (every 15 mins) in three representative transport zones in the metropolitan area of Madrid.

Another way of analysing the variability in dynamic accessibility according to transport zones is to map the differences between the assessment scenarios and the reference scenario in each zone at several times of day. In the reference scenario (Figure 4) the spatial distribution of accessibility shows a characteristic centre-suburban pattern. The highest accessibility is recorded in the central transport zones with a high land-use intensity (concentration of activities and/or population), whereas in the suburbs the accessibility values are much lower. However, the spatial distribution varies throughout the day. To analyse the dynamic scenarios in detail, three times were selected, corresponding to the morning peak time, the morning valley, and the afternoon peak time. To observe the differences more clearly, each of the dynamic scenarios is compared to the reference scenario at those three times (Figure 5). Additionally, three videos have been produced showing these differences throughout the full day (videos 3 to 5).

The comparison between the dynamic congestion scenario and the reference scenario (Figure 5a and video 3) shows very marked differences for the journeys that reach their destination at 8:30 am (maximum congestion), less marked at 5:30 pm (start of the afternoon peak time) and practically non-existent at midday (average congestion). These results are consistent with the curves in Figure 1 and with Tables 2 and 3, but they raise some further considerations that are worth exploring. The transport zones in the south, east and west register the greatest negative differences in accessibility in the morning peak, whereas the situation improves in the centre and north, which concentrate a large proportion of employment, and where there is therefore a predominance of inbound rather than outbound journeys at that time of day. Practically the opposite situation occurs at the start of the afternoon peak time: the greatest negative differences correspond to the centre and particularly the north. The flows are now reversed compared to the morning, and these areas that concentrate a large volume of employment now begin to register high congestion in outbound journeys.

The comparison between the dynamic attractiveness scenario and the reference scenario (Figure 5b and video 4) reveals significant differences at 8:30 am, but in the positive sense. The increase in population density in the centre and north causes an increase in accessibility in these areas, which contain a high concentration of jobs.

The dynamic accessibility scenario is the result of the joint effect of the changes in network congestion and the attractiveness of the destinations, with a much greater influence of the first than the second (Figure 5c and video 5). In fact, Figures 5a and 5c are very similar. The most important difference is the reinforcement of the effects of network performance and population concentration in the city centre at the morning peak time. The general resulting sequence can be seen in the vide 5 and reveals how the southern, eastern and western zones are the most negatively affected at the morning peak time, whereas the areas in the centre and north are most adversely affected during the afternoon peak, essentially due to the effects of congestion.

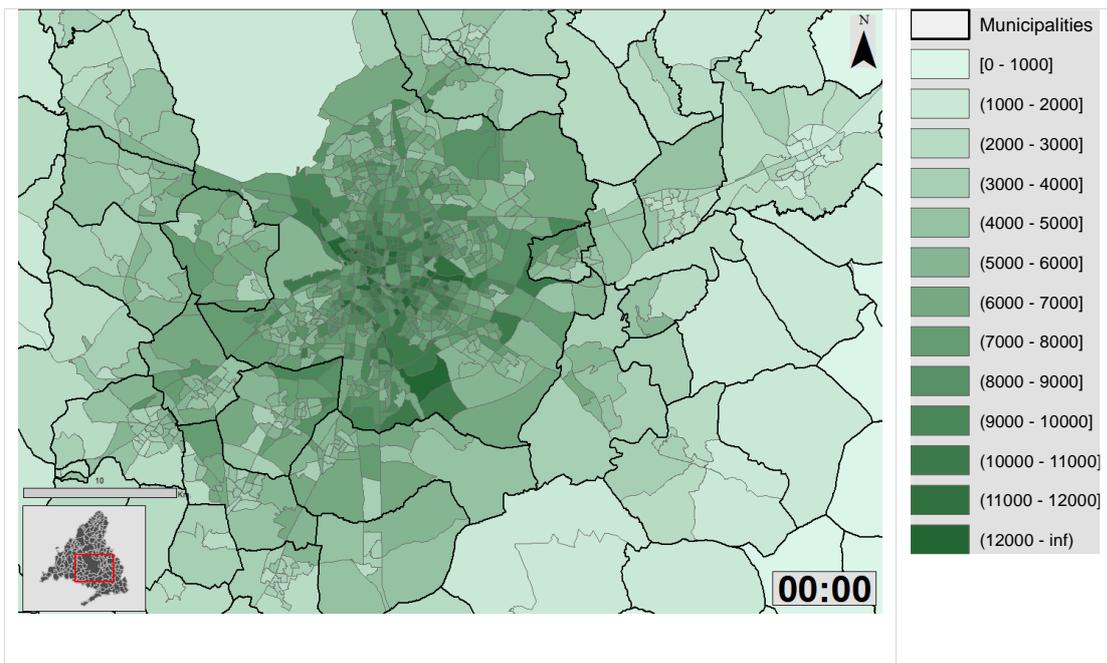

Figure 4: Spatial distribution of accessibility in the reference scenario (network congestion and destination attractiveness remain static)

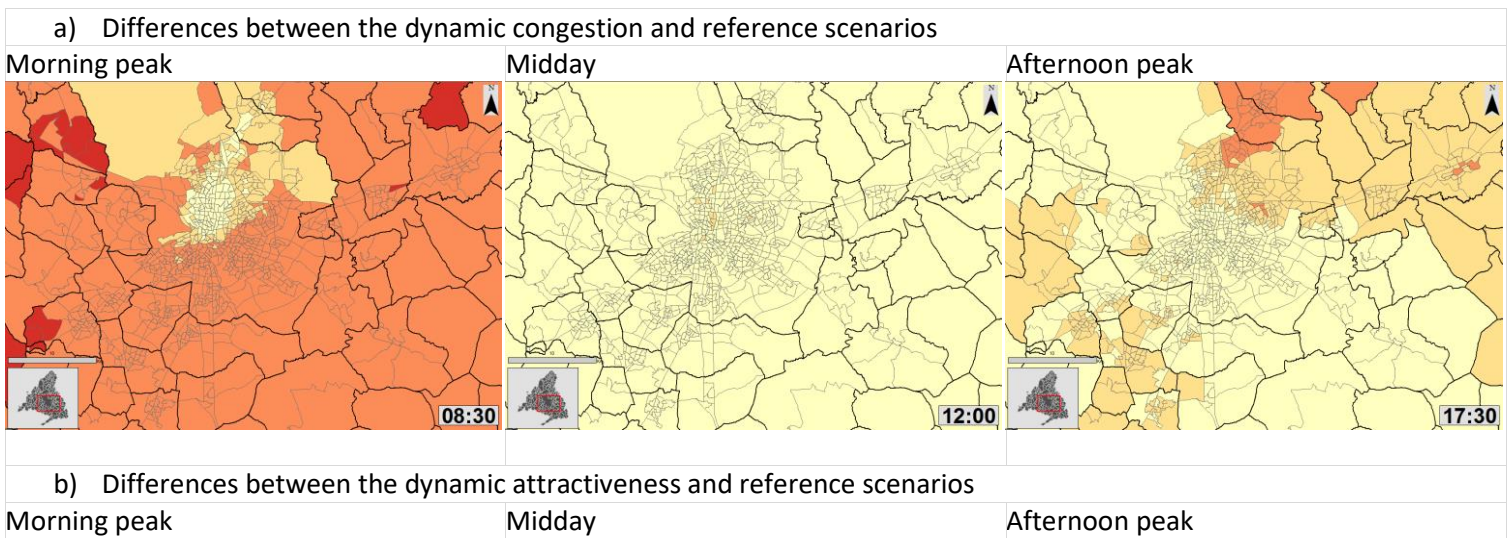

a) Differences between the dynamic congestion and reference scenarios

| Morning peak | Midday | Afternoon peak |

b) Differences between the dynamic attractiveness and reference scenarios

| Morning peak | Midday | Afternoon peak |

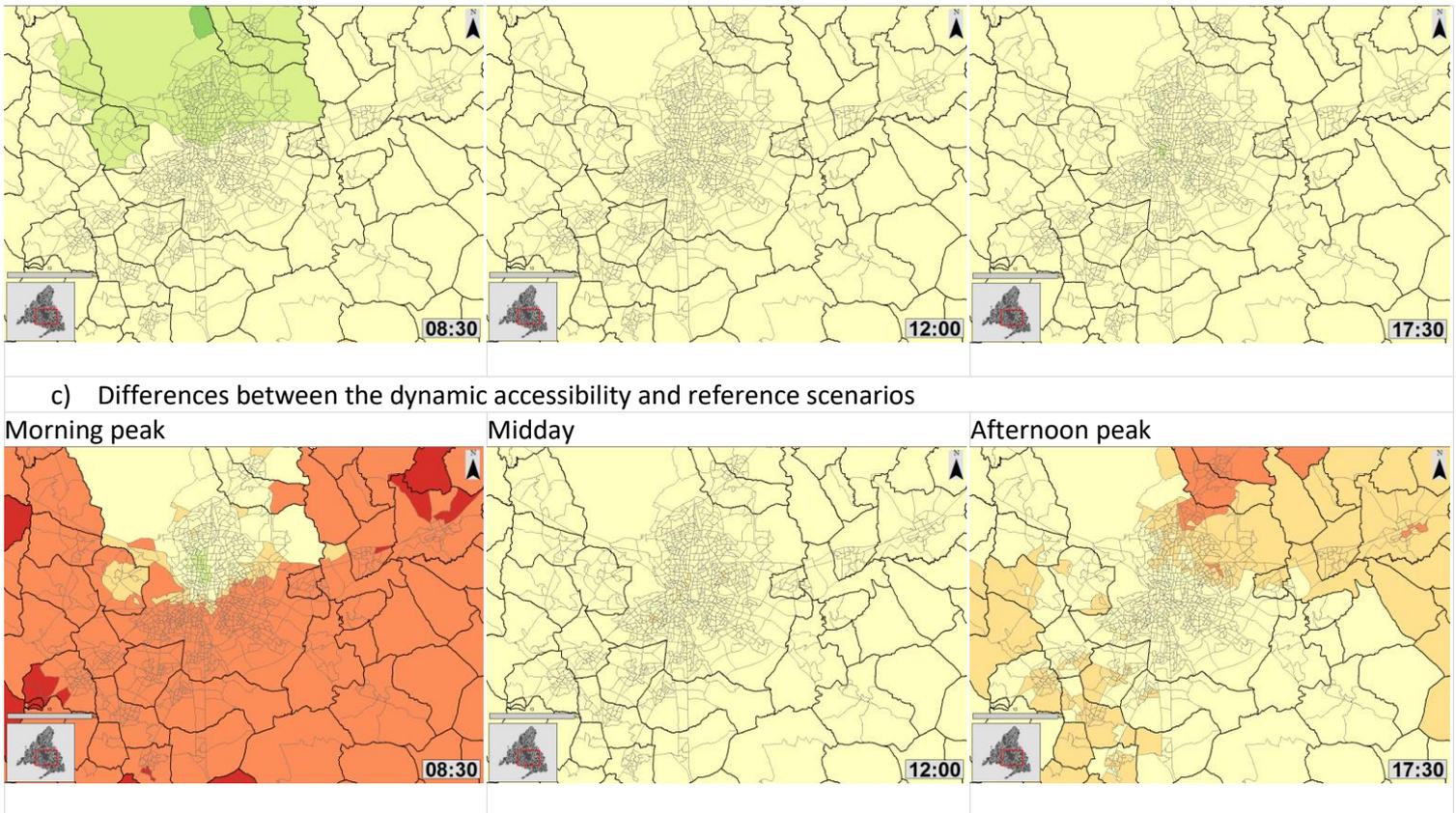

c) Differences between the dynamic accessibility and reference scenarios

Figure 5: Differences between the dynamic scenarios and the reference scenario

Finally, the coefficient of variation in accessibility between transport zones and within each transport zone was calculated in each scenario. The aim of the first is to analyse the spatial variability between transport zones and identify the times and scenarios with the most homogeneous distribution, while the second identifies the temporal variability in accessibility in each transport zone in order to differentiate the transport zones that have a greater temporal stability from those that have greater variability.

Tables 2 and 3 show the descriptive statistics for accessibility in the static and dynamic scenarios respectively. To facilitate comparison between the dynamic scenarios and the reference scenario, the differences were calculated between the values in each scenario and the reference scenario and expressed in the form of ratios. At the morning peak time the discrepancies in accessibility in both congestion and population distribution increase compared to the reference scenario (ratios of 1.09 and 1.03, respectively), producing a 9% and 3% higher coefficient of variation (variability between transport zones) in these dynamic scenarios than in the reference scenario. Figure 5 shows that the more suburban areas with poorer accessibility suffer greater effects of congestion at this time, while the central areas (centre-north) benefit from an increase in population. Both effects are self-reinforcing and lead to a considerable increase in inequalities in the dynamic accessibility scenario compared to the reference scenario (ratio of 1:13).

At the start of the afternoon peak time the increase in the discrepancies between the dynamic accessibility scenario and the reference scenario is weak (ratio of 1:03), owing to the concentration of activities (ratio of 1.03) rather than to congestion (ratio of 1.00), whereas at midday the reduction in congestion (ratio 0.98) also produces a decrease in the discrepancies in dynamic accessibility (0.98).

Table 2: Descriptive statistics of accessibility in the reference scenario

|  | Accessibility values |
|---|---|
| Number of transport zones | 1,010 |
| Min | 325.42 |
| Max | 13,840.83 |
| Mean | 6,771.15 |
| SD | 2,772.49 |
| CV | 40.95 |

Table 3. Descriptive statistics of accessibility in the dynamic scenarios

| Dynamic congestion scenario | | | | | | |
|---|---|---|---|---|---|---|
| | Accessibility | | | Differences (dynamic congestion scenario/reference scenario) | | |
| | Morning peak | Midday | Afternoon peak | Morning peak | Midday | Afternoon peak |
| Min | 276.74 | 325.21 | 316.93 | 0.85 | 1.00 | 0.97 |
| Max | 12 876.25 | 13 763.07 | 13 321.02 | 0.93 | 0.99 | 0.96 |
| Mean | 6 096.54 | 6 648.37 | 6 387.61 | 0.90 | 0.98 | 0.94 |
| SD | 2 732.87 | 2 669.58 | 2 618.75 | 0.99 | 0.96 | 0.94 |
| CV | 44.83 | 40.15 | 41.00 | 1.09 | 0.98 | 1.00 |
| Dynamic attractiveness scenario | | | | | | |
| | Accessibility | | | Differences (dynamic attractiveness scenario/reference scenario) | | |
| | Morning peak | Midday | Afternoon peak | Morning peak | Midday | Afternoon peak |
| Min | 303.20 | 321.25 | 322.71 | 0.93 | 0.99 | 0.99 |
| Max | 14 201.09 | 13 698.87 | 14 252.53 | 1.03 | 0.99 | 1.03 |
| Mean | 6 967.31 | 6 765.65 | 6 894.79 | 1.03 | 1.00 | 1.02 |
| SD | 2 929.16 | 2 758.65 | 2 912.05 | 1.06 | 1.00 | 1.05 |
| CV | 42.04 | 40.77 | 42.24 | 1.03 | 1.00 | 1.03 |
| Dynamic accessibility scenario | | | | | | |
| | Accessibility | | | Differences (dynamic accessibility scenario/reference scenario) | | |
| | Morning peak | Midday | Afternoon peak | Morning peak | Midday | Afternoon peak |
| Min | 253.70 | 321.07 | 314.28 | 0.78 | 0.99 | 0.97 |
| Max | 13 227.94 | 13 622.12 | 13 592.18 | 0.96 | 0.98 | 0.98 |
| Mean | 6 235.76 | 6 643.06 | 6 504.96 | 0.92 | 0.98 | 0.96 |
| SD | 2 892.88 | 2 654.65 | 2 753.08 | 1.04 | 0.96 | 0.99 |
| CV | 46.39 | 39.96 | 42.32 | 1.13 | 0.98 | 1.03 |

Figure 6 shows the temporal variability in accessibility in each transport zone in the dynamic scenarios. In the dynamic accessibility scenario (Figure 6c), the centre, and the north to a minor degree, present the lowest variation coefficients. Once again this distribution is a result of the interplay between the network congestion dynamic and the destination attractiveness dynamic. Congestion generates more uneven accessibility profiles (Figure 6a), and has a greater effect on the more suburban areas. The population density dynamic has less effect on the daily variation in accessibility in each zone (Figure 6b), but more in the centre and north of the city.

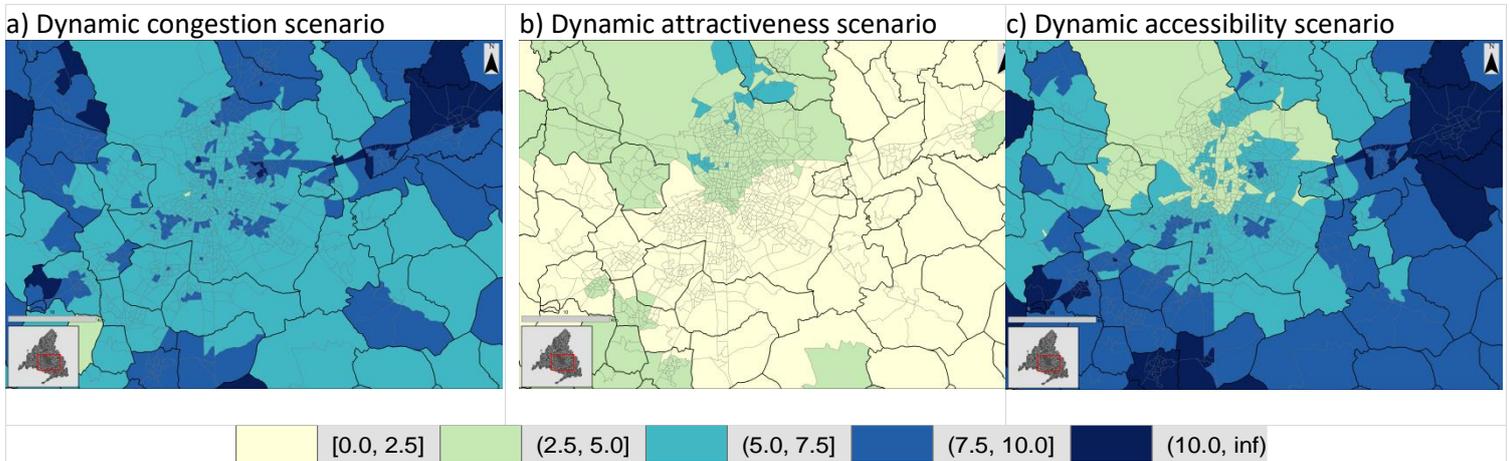

**Figure 6: Coefficient of variation in accessibility distribution profiles (every 15 minutes) according to transport zones and scenarios**

**6 CONCLUSIONS**

The new data sources offer new possibilities for the dynamic analysis of accessibility. The recent availability of traffic data from commercial navigation companies (TomTom, Nokia Here, Garmin), and as (limited) open data (Google), is a genuine leap forwards for the study of dynamic travel times. In addition, data on people's location during the day is becoming more spatially and temporally disaggregated, thus contributing to greater detail in the land-use aspect of accessibility measures. For example, geolocated tweets can be downloaded free of charge and used to map the intensity of use of each part of the city during the day.

In this research we have used the increasing availability of big data to overcome the temporal restrictions of previous studies at the urban scale. We consider the two components of accessibility dynamically: travel times to reach the destinations, and the attractiveness of the destinations themselves. For the first we use data from the TomTom company, which provides travel times for each section of the network in 15-minute intervals. For the second we use the number of active tweeters in each transport zone in 15 minute intervals as a proxy. These data highlight the variation in the spatial distribution of the population depending on the time of day, and show that it tends to be concentrated in the centre in the morning and afternoon and in residential areas in the suburbs at night.

The results of the dynamic accessibility analyses reveal that in general the poorest accessibility conditions are recorded at the morning and afternoon peak times due to the increase in congestion, although its effects are partially offset by the distribution of the population density, which increases in the city centre at these times. As expected, the best accessibility conditions occur at night in a situation of free flow, although the population tends to be dispersed throughout the suburbs.

These general results conceal marked contrasts between transport areas. The distribution of accessibility in the reference scenario shows a typical accessibility gradient between the centre and the suburbs in an average congestion scenario and an average distribution of population densities. However, these accessibility conditions change significantly depending on the time of day. At the morning peak time the greatest decrease in accessibility occurs in suburban residential zones, as the predominance of inbound travel causes serious problems of congestion, while the population tends to be concentrated in the centre. In contrast, at the afternoon peak time the transport zones in the centre and north, which contain the highest proportion of jobs, are the most negatively affected by the effects of congestion, at a time when outbound journeys predominate. The calculation of the coefficients of variation of the profiles of each transport zone shows that in general the transport zones in the suburbs register a greater temporal variation in their accessibility conditions.

Dynamics of accessibility throughout the day give policymakers greater insight into accessibility issues that are otherwise masked in static accessibility analyses. Accessibility conditions change throughout the day, as do the causes of these changes, namely congestion in the road network and the population distribution in the city. Each transport zone has its own accessibility profile, and thus its own specific problems, which require solutions that are also specific.

This research also represents a further step forward for accessibility analyses by considering the two accessibility components dynamically and determining which one has the greatest impact in each transport zone and at each time of day. However, the work has certain limitations that can be overcome in future research. The analysis of geolocated tweets shows that, as expected, the density of tweeters varies considerably throughout the day, increasing in areas of activity during the daytime and in residential areas at night. However, Twitter is a fairly biased source and is possibly not the most effective for use as a proxy for the variation in population density throughout the day. In future work we will use data from more reliable sources than Twitter –specifically mobile phone data– to consider the time variation in the available activities. The analysis of dynamic accessibility in public transport is beyond the scope of this research, but it will be examined in the future using data from Google Transit and mobile phones.

## ACKNOWLEDGMENTS

The authors gratefully acknowledge funding from the ICT Theme of the European Union's Seventh Framework Programme (INSIGHT project - Innovative Policy Modelling and Governance Tools for Sustainable Post-Crisis Urban Development, GA 611307), the Spanish Ministry of Economy and Competitiveness (TRA2015-65283-R and FPDI 2013/17001), and the Madrid Regional Government (SOCIALBIGDATA-CM, S2015/HUM-3427).